\documentclass[10pt,preprint]{aastex}

\begin{document}

\shortauthors{Luhman \& Steeghs}
\shorttitle{Spectroscopy in $\eta$ Cha and MBM~12}

\title{Spectroscopy of Candidate Members of the $\eta$ Cha and MBM~12 
Young Associations \altaffilmark{1}}

\author{K. L. Luhman \& D. Steeghs}
\affil{Harvard-Smithsonian Center for Astrophysics, 60 Garden St.,
Cambridge, MA 02138}

\email{kluhman, dsteeghs@cfa.harvard.edu}

\altaffiltext{1}{Based on observations performed at Las Campanas Observatory
and the MMT Observatory.
This publication makes use of data products from the Two Micron All  
Sky Survey and the Deep Near-Infrared Survey of the Southern Sky.}

\begin{abstract}

We present an analysis of candidate members of the $\eta$~Cha and MBM~12A
young associations.
For an area of 0.7~deg$^2$ toward $\eta$~Cha, we have performed a search for
members of the association by combining $JHK_s$ photometry from 2MASS and
$i$ photometry from DENIS 
with followup optical spectroscopy at Magellan Observatory.
We report the discovery of three new members with spectral types of 
M5.25-M5.75, corresponding to masses of 0.13-0.08~$M_{\odot}$ by theoretical
evolutionary models.
Two and three of these members were found independently 
by Lyo and coworkers and Song and coworkers, respectively. 
Meanwhile, no brown dwarfs were detected in $\eta$~Cha down to the 
completeness limit of 0.015~$M_{\odot}$. 
For MBM~12A, we have obtained spectra of three of the remaining candidate 
members that lacked spectroscopy at the end of the survey by Luhman,
all of which are found to be field M dwarfs.
Ogura and coworkers have recently presented four ``probable" members of 
MBM~12A. However, two of these objects were previously classified 
as field dwarfs by the spectroscopy of Luhman. In this work, we find that
the other two objects are field dwarfs as well.

\end{abstract}


\section{Introduction}

The environs of the B8 star $\eta$ Cha and the dark cloud MBM~12 have 
been revealed as sites of small associations of newly formed stars.
Through deep {\it ROSAT} observations of 0.35~deg$^2$ toward $\eta$ Cha,
\citet{mam99} discovered X-ray emission from 12 sources, including $\eta$ Cha
itself. From {\it Hipparcos} parallactic distances and proper motions,
the two earliest stellar counterparts and an early-type star without X-ray
emission were found to be comoving at a distance of $\sim100$~pc.
\citet{mam99} confirmed the youth of the remaining 10 late-type stars through 
measurements of H$\alpha$ emission and Li absorption, which indicated their
probable membership in an association with the early-type stars. 
An age of 4-10 Myr has been inferred for these stars through 
comparisons of their positions on the Hertzsprung-Russell (H-R) diagram 
to theoretical isochrones \citep{mam99,law01}.
Recently, five additional low-mass members of the association 
(0.08-0.3~$M_{\odot}$) have been identified through spectroscopy of candidates 
appearing in color-magnitude diagrams generated from photometry at $V$ and $I$
\citep{law02,lyo04} and from data in the USNO-B1 and Two-Micron 
All-Sky Survey (2MASS) catalogs \citep{sz04}.
Although not precisely determined, the completeness limits of these latest 
surveys appeared to be near 0.1~$M_{\odot}$. 

Surveys for sources of H$\alpha$ and X-ray emission toward 
cloud 12 from \citet{mbm85} (MBM~12) have resulted in the discovery of seven 
T Tauri stars (e.g., \citet{hb88}). 
MBM~12 was long believed to be the nearest molecular cloud at a distance 
of 50-100 pc \citep{hbm86,hea00}, making the associated group of young 
stars (MBM~12A) a particularly attractive site for studies of star formation.
However, during a search for new members, \citet{luh01} noticed that the 
members of MBM~12A exhibited anomalously old ages ($\sim100$~Myr) 
when placed on the H-R diagram with the published distances.
Through new distance estimates in that work and subsequent studies 
\citep{str02,and02}, the distance of MBM~12 cloud is now firmly established 
near a value of $\sim300$~pc.
Using this distance, theoretical isochrones produce an age of $\sim2$~Myr, which
is consistent with the evolutionary state of the association implied by 
various signatures of newly formed stars \citep{luh01}.
The magnitude-limited survey by \citet{luh01} achieved a completeness limit
of 0.03~$M_{\odot}$ and uncovered five new members (0.1-0.4~$M_{\odot}$),
bringing the total known membership to a dozen sources at a resolution of 
$\sim1\arcsec$. \citet{ogu03} have since identified four ``probable" members of 
MBM~12A via H$\alpha$ emission detected with slitless grism spectroscopy.

In this paper, we present spectroscopy of candidate members of the $\eta$~Cha 
and MBM~12 associations. 
For $\eta$~Cha, we select candidate members of the
association by combining $i$ photometry from the Deep 
Near-Infrared Survey of the Southern Sky (DENIS) and $JHK_s$ data from 
the Two-Micron All-Sky Survey (2MASS) (\S~\ref{sec:ident1}), 
measure their spectral types and determine
their status as field stars or association members (\S~\ref{sec:class1}), 
evaluate the completeness of the survey (\S~\ref{sec:complete}), and place
the known members of $\eta$~Cha on the H-R diagram (\S~\ref{sec:hr}).
For MBM~12A, we select candidate members from \citet{luh01} and 
\citet{ogu03} for spectroscopy (\S~\ref{sec:ident2}) and use these data to 
measure spectral types and assess membership (\S~\ref{sec:class2}).

\section{$\eta$ Cha}

\subsection{Selection of Candidate Members}
\label{sec:ident1}

\begin{figure}
\plotone{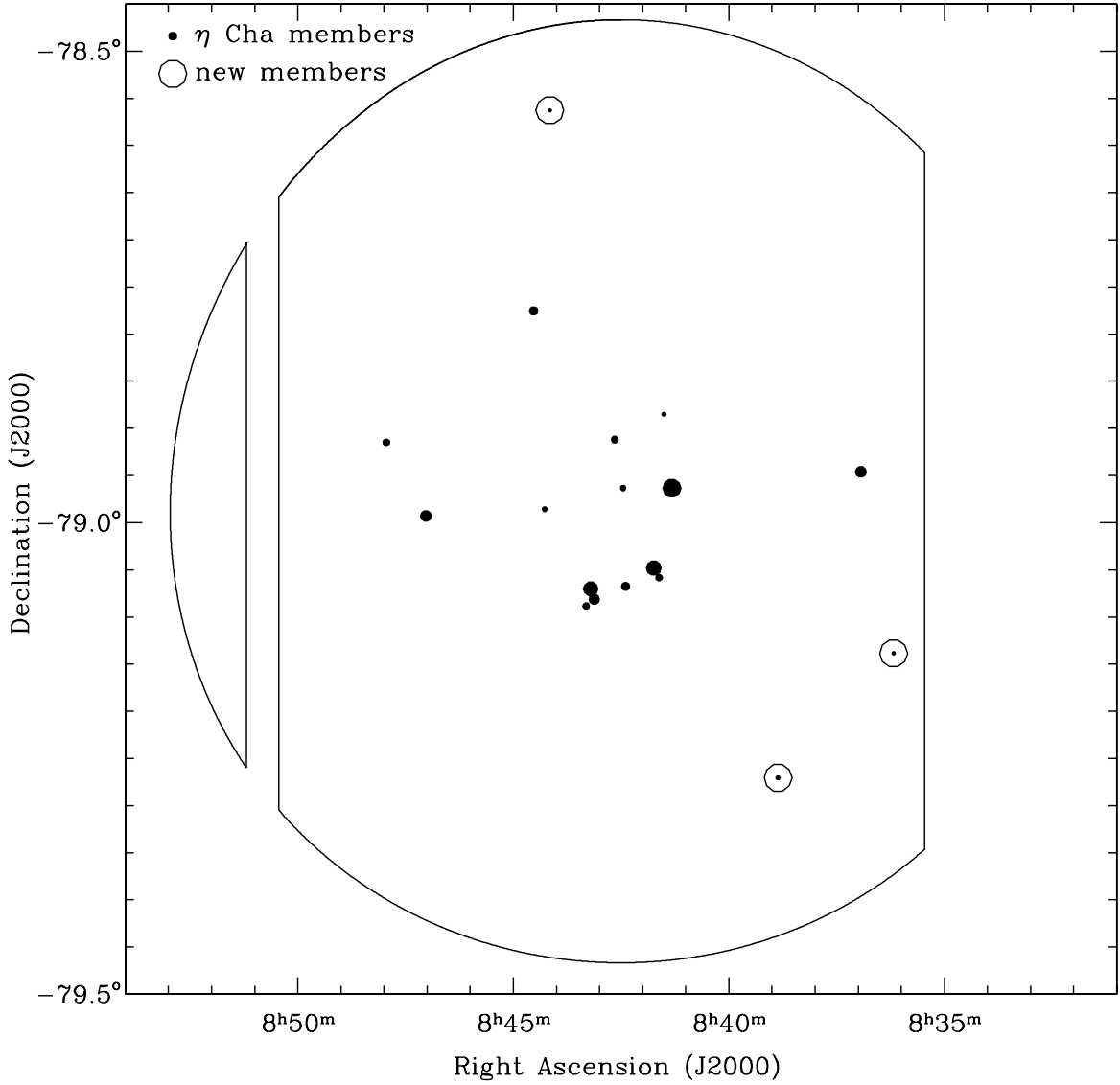}
\caption{
Map of the 18 known members of the $\eta$~Cha young association. 
The masses of these sources range from 0.08 to 3.2~$M_\odot$ and are 
represented by the sizes of the points.
The three circled members were discovered in this work and independently by 
\citet{lyo04} and \citet{sz04}. The boundaries of our survey field are 
indicated.
}
\label{fig:map}
\end{figure}

\begin{figure}
\plotone{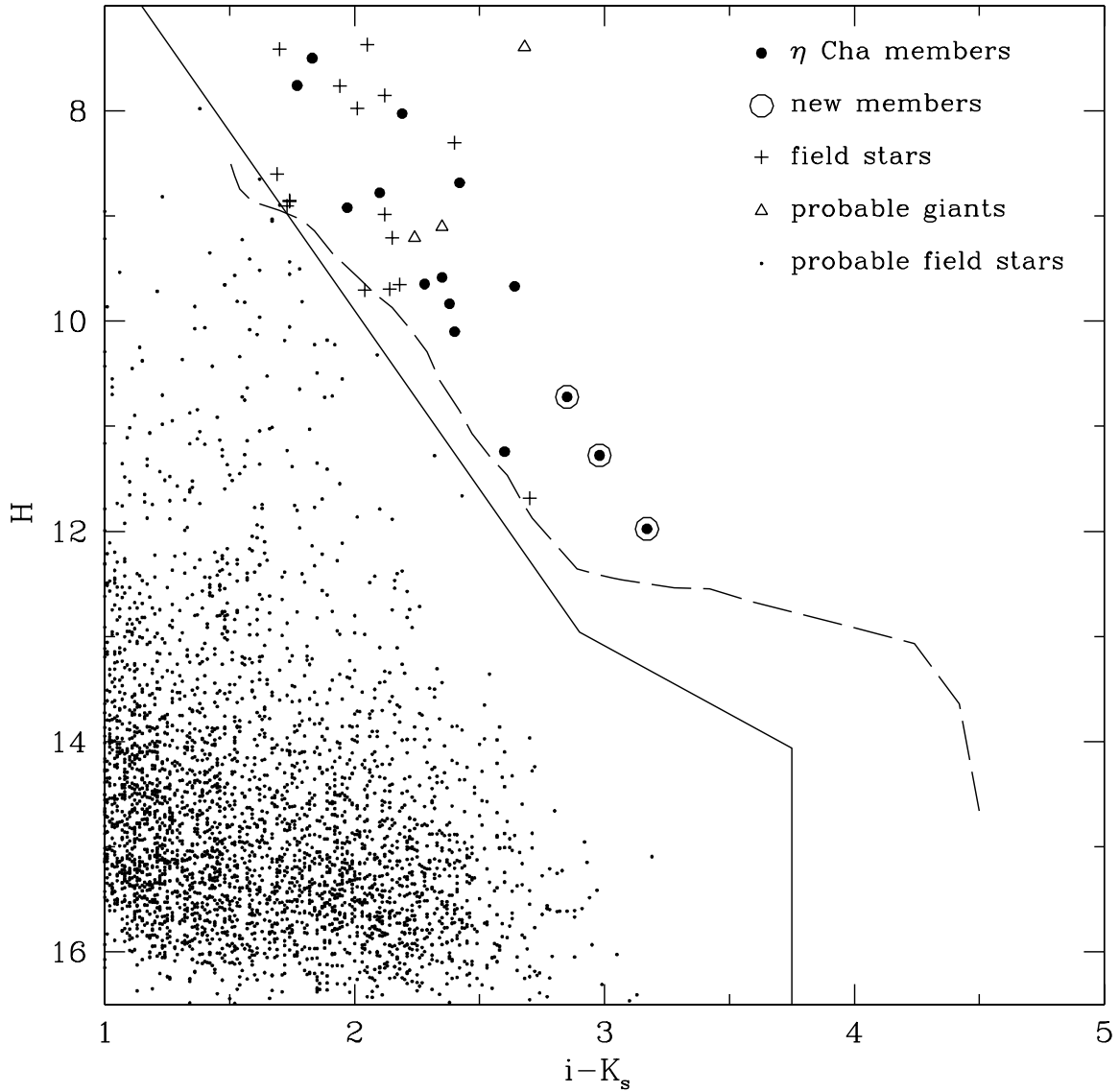}
\caption{
Color-magnitude diagram for the field toward the $\eta$~Cha young association
in Figure~\ref{fig:map}.
The 15 known late-type members of the association are shown 
({\it large points}). We also indicate sources spectroscopically classified
as new members ({\it circles}) and as field stars ({\it plusses}).
The dashed line is the 10~Myr isochrone (1-0.015~$M_{\odot}$) from the 
evolutionary models of \citet{bar98}. 
Objects below the solid boundary are probable field stars ({\it small points}). 
The three remaining sources above this boundary without spectroscopy
have positions in Figure~\ref{fig:jhhk} that are indicative of field giants
({\it open triangles}).
These measurements are from DENIS ($i$) and 2MASS ($H$, $K_s$).}
\label{fig:ik}
\end{figure}

\begin{figure}
\plotone{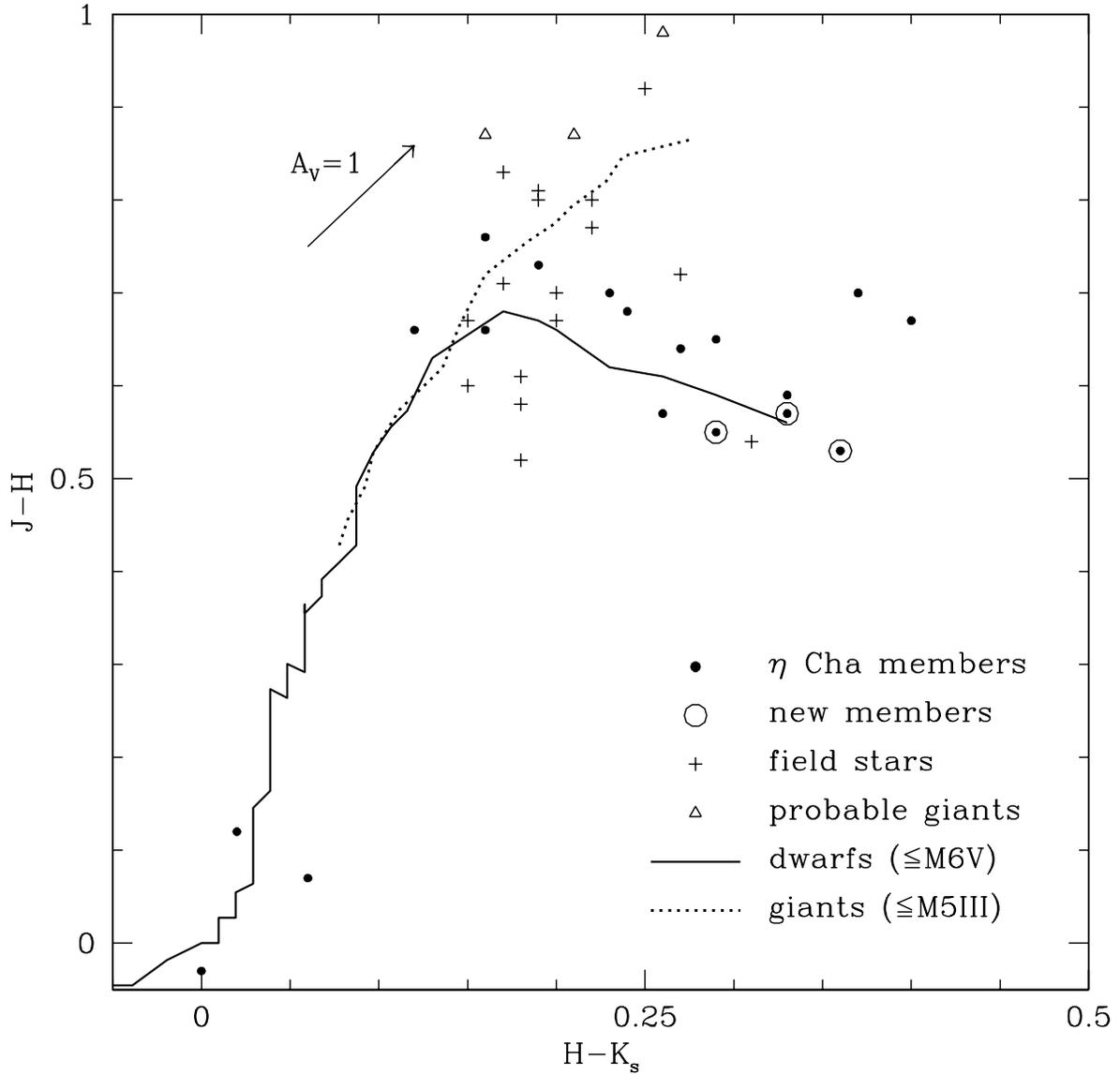}
\caption{
$H-K_s$ versus $J-H$ for the field toward the $\eta$~Cha young association
in Figure~\ref{fig:map}.
The symbols are the same as in Figure~\ref{fig:ik}, with the addition
of sequences for typical field dwarfs ({\it solid line}; $\leq$M6V) 
and giants ({\it dotted line}; $\leq$M5~III). Objects identified as
probable field stars by their location below the boundary in 
Figure~\ref{fig:ik} are omitted. These measurements are from 2MASS.}
\label{fig:jhhk}
\end{figure}

To search for new members of a stellar group, one can select stars
that exhibit the colors and magnitudes expected of members in color-color
and color-magnitude diagrams and obtain followup spectroscopy to distinguish
between field stars and bonafide members. For our survey of $\eta$~Cha, 
we utilized $i$ photometry from the DENIS Second Release and $JHK_s$ 
photometry from the 2MASS Point Source Catalog. 
These surveys exhibit a signal-to-noise of 10 at $i\sim17$, $J\sim15.8$, 
$H\sim15.1$, and $K_s\sim14.3$.
We considered all sources
appearing in both databases within a radius of $0\fdg5$ from 
$\alpha=8^{\rm h}42^{\rm m}30^{\rm s}$, $\delta=-78\arcdeg58\arcmin00\arcsec$
(J2000). The size of this survey field was selected to extend well beyond 
the positions of the known members of the association.
Two areas within this radius were not available from the DENIS 
Second Release. They are defined approximately by 
$\alpha=8^{\rm h}50^{\rm m}27^{\rm s}$ to $8^{\rm h}51^{\rm m}11^{\rm s}$
and $\alpha<8^{\rm h}35^{\rm m}27^{\rm s}$.
The resulting survey field has a size of 0.7~deg$^2$ and is illustrated in
Figure~\ref{fig:map}.
These data are shown on the diagram of $i-K_s$ versus $H$ in 
Figure~\ref{fig:ik}.
We also include the 10~Myr isochrone from \citet{bar98} for masses
of 0.015 to 1~$M_{\odot}$ at a distance of 97~pc \citep{mam99}. 
This isochrone was converted to photometric magnitudes from predicted
effective temperatures and bolometric luminosities in the 
manner described by \citet{luh03a}.
To separate candidate members of $\eta$~Cha from likely field stars, we
defined a boundary that is below this isochrone. To account for the larger
photometric errors that would be exhibited by any members at the faintest 
levels, we placed the boundary at a larger distance below the isochrone at 
lower masses.
Above this boundary, there are 22 stars that are not previously known  
members. We selected 16 of these candidates for spectroscopy.
Among the six remaining sources, three stars were already 
classified as field giants by \citet{law02}. The other three candidates, 
2MASS~J08501189-7844236, J08382996-7924507, and J08501972-7906344, were not
observed because they are likely to be field giants according to their
$J-H$ and $H-K_s$ colors, as shown in Figure~\ref{fig:jhhk}.  
The 12 known members at K and M spectral types were
included in our spectroscopic sample as well. 
Table~\ref{tab:log1} summarizes long-slit spectroscopy of these 28 targets.
The observing and data reduction procedures are the same as those
described by \citet{luh04}.

\subsection{Classification of Candidate Members}
\label{sec:class1}

\begin{figure} 
\vspace{-4cm}
\plotone{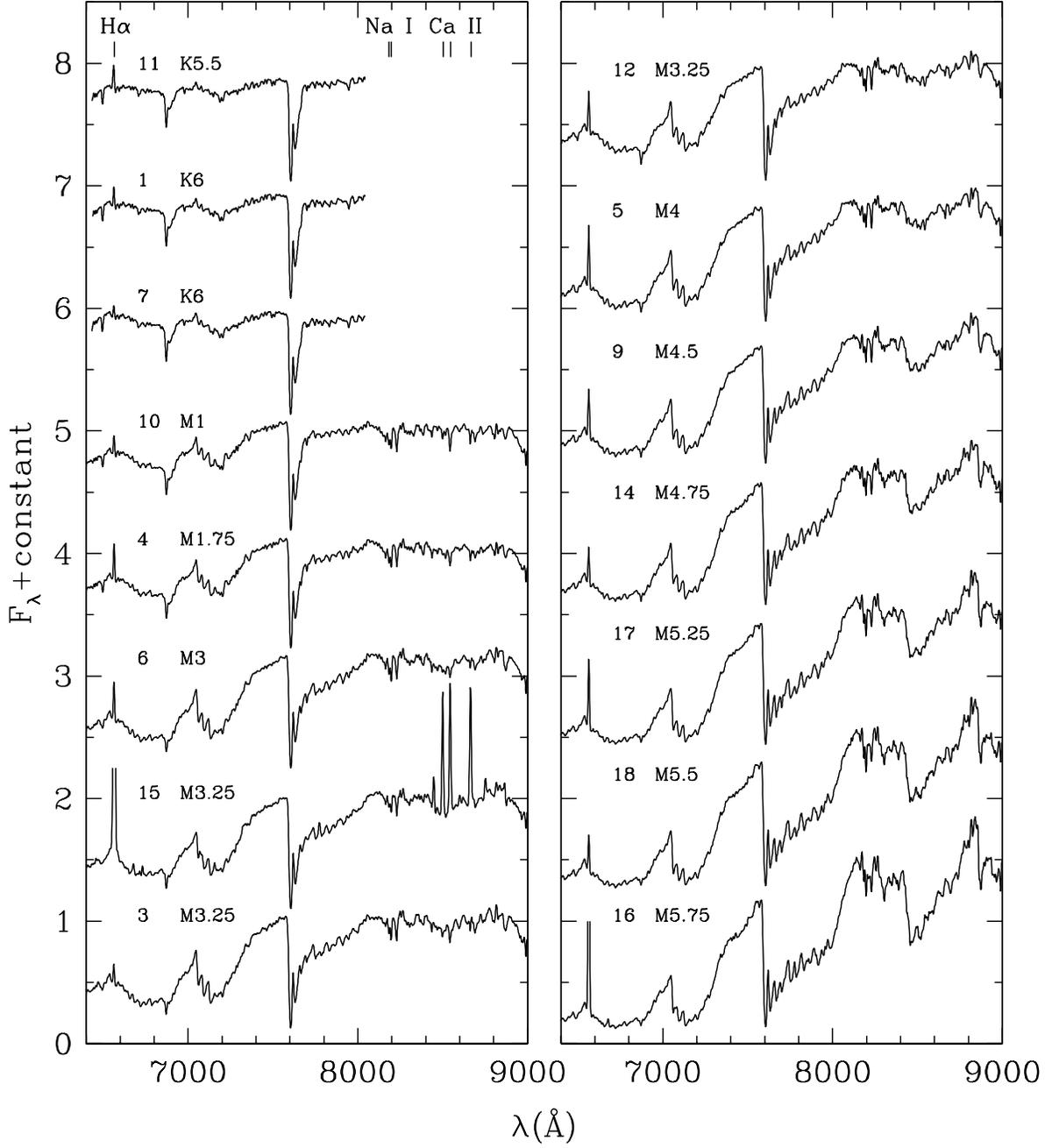}
\caption{
Low-resolution spectra of the 15 known late-type members of the $\eta$ Cha 
young association. Source identifications are from Table~\ref{tab:mem}.
The data are displayed at a resolution of 8~\AA\ and are 
normalized at 7500~\AA.}
\label{fig:ec}
\end{figure}

We measured spectral types and assessed membership for the 16 candidate 
members of $\eta$~Cha by applying the methods of classification described 
by \citet{luh04} for a similar set of data in Chamaeleon~I.
In this way, we classified 13 candidates as field dwarfs and background giants. 
Astrometry, photometry, and spectral types for these sources are listed in 
Table~\ref{tab:back}. We also include the three field stars identified 
spectroscopically by \citet{law02}. 
Among the three remaining candidates, source 16 has 
emission in H$\alpha$ ($W_\lambda=77$~\AA) that is
above the levels observed for active field dwarfs \citep{giz00,giz02}
and emission in He~I at 6678~\AA. 
Although emission lines of this kind can arise from flaring field dwarfs,
the low duty cycle of such flares \citep{giz00} combined
with the presence of these strong lines during the separate observations 
in this work and in \citet{sz04} indicates that the emission source 
is probably a young star rather than a flaring field dwarf.
The spectra of this object and the other two remaining candidates exhibit 
conclusive evidence of youth, and thus membership in $\eta$~Cha, in the form 
of the weak K~I and Na~I absorption features that are characteristic of 
pre-main-sequence objects \citep{mar96,luh99}. 
These three new members were independently discovered in the survey by 
\citet{sz04}. Sources 17 and 18 were also reported by \citet{lyo04}. 
The spectra of the three new members and the 12 previously known late-type 
members are presented in Figure~\ref{fig:ec}.
Astrometry, photometry, spectral types, and evidence of membership for the
18 known members of $\eta$~Cha are compiled in Table~\ref{tab:mem}.
The $I$ photometry from \citet{law01}, \citet{law02}, and \citet{lyo04}
agree with the $i$ data from DENIS for most of the members of $\eta$~Cha
with the exception of the third brightest member, source 13 (HD~75505),
for which $i$ is fainter than $I$. 
The two brightest stars, sources 2 and 8 ($\eta$~Cha and RS~Cha) were not 
measured at $I$, but their $i$ data are also much fainter than expected 
relative to the IR photometry from 2MASS. Therefore, the DENIS $i$
measurements for these three members are not included in Table~\ref{tab:mem}.

\subsection{Completeness of Survey}
\label{sec:complete}

\begin{figure}
\plotone{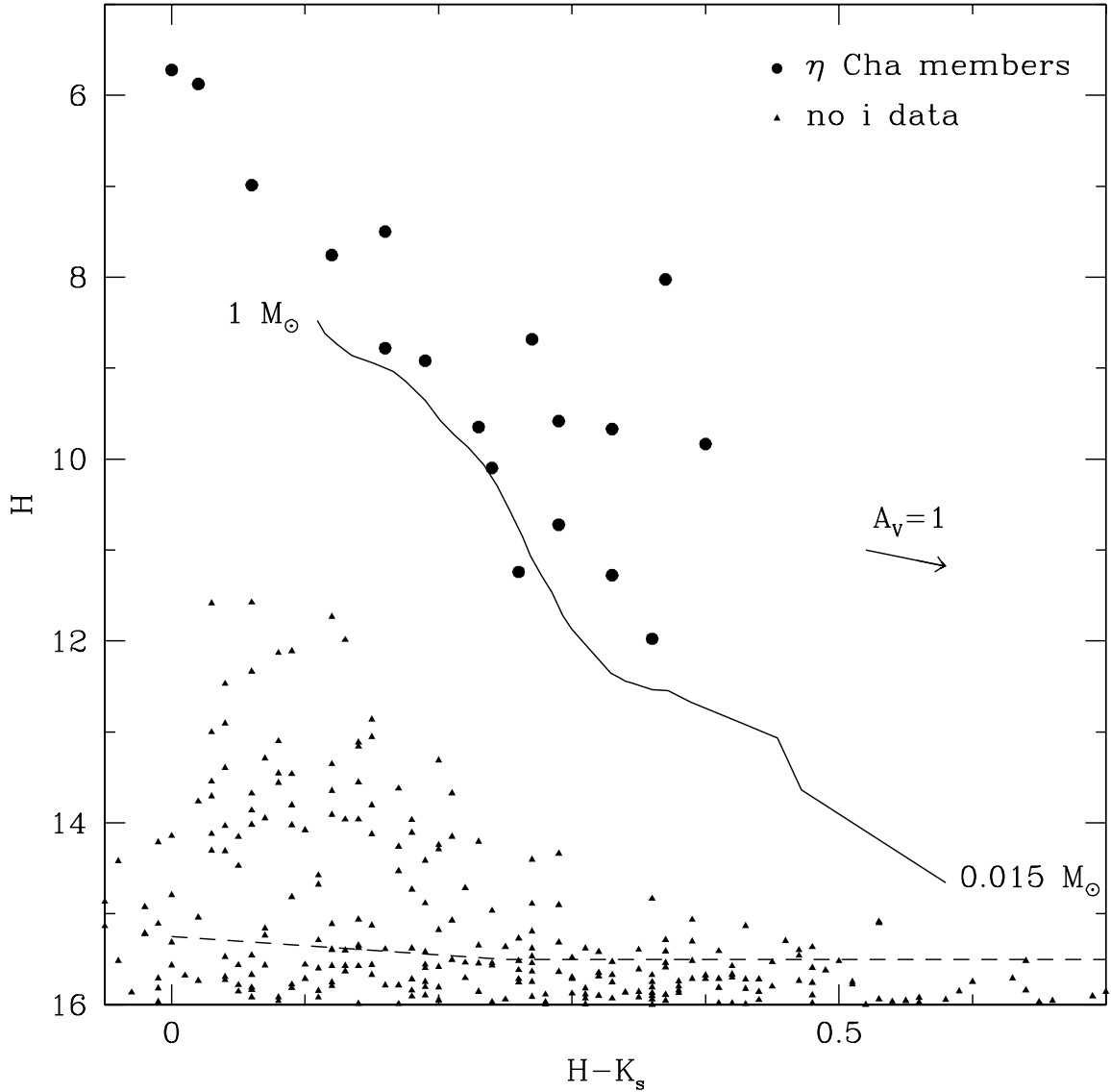}
\caption{
$H-K_s$ versus $H$ for the field toward the $\eta$~Cha young association
in Figure~\ref{fig:map}.
Known members of the association ({\it points}) and sources without 
detections at $i$ by DENIS ({\it triangles}) are indicated.
We have omitted the field stars identified through spectroscopy 
(Tables~\ref{tab:back}) as well as objects that are probable field stars
by their location below the boundary in Figure~\ref{fig:ik}.
The solid line is the 10~Myr isochrone 
from the evolutionary models of \citet{bar98}. These measurements are from 2MASS
and have completeness limits of $H=15.5$ and $K_s=15.25$ ({\it dashed line}).}
\label{fig:hk}
\end{figure}

Because low-mass, cool sources are most easily detected at near-IR wavelengths, 
we use the diagram of $H-K_s$ versus $H$ in Figure~\ref{fig:hk} to evaluate 
the mass completeness of our survey for new members of $\eta$~Cha.
The completeness limits of the 2MASS photometry are taken to be the magnitudes 
at which the logarithm of the number of sources as a function of magnitude 
departs from a linear slope and begins to turn over 
($H\sim15.5$, $K_s\sim15.25$).
Without quantitative support, \citet{sz04} asserted that late-M members of 
$\eta$~Cha are not detectable with data from 2MASS.
However, according to the evolutionary models of \citet{bar98}, the 
completeness limits of the 2MASS data correspond to a mass of 
$\sim0.01$~$M_{\odot}$ and a spectral type of $\sim$L0 for objects at 
the age and distance of the association (6~Myr, 97~pc).

In Figure~\ref{fig:hk}, we have plotted the theoretical isochrone from
\citet{bar98} for 10~Myr, which is near the upper limit we find for
the age of the $\eta$~Cha association (\S~\ref{sec:hr}). 
We have omitted the field stars identified with spectroscopy as well as 
objects that are probable field stars by their location below the 
boundary in Figure~\ref{fig:ik}. The three stars lacking spectroscopy
that were identified as likely field giants by Figure~\ref{fig:jhhk} are also
excluded. The remaining IR sources consist of known members of the
association and objects that were not detected at $i$ by DENIS.
As demonstrated in Figure~\ref{fig:hk}, our search for new members is complete 
to masses of 0.015~$M_{\odot}$, which is near the deuterium burning mass limit 
\citep{bur97}.

\subsection{H-R Diagram}
\label{sec:hr}

In this section, we estimate effective temperatures and bolometric 
luminosities for the members of $\eta$~Cha, place these data on the 
H-R diagram, and use theoretical evolutionary models to infer masses and ages.

In the following analysis, standard dwarf colors are taken from the compilation
of \citet{kh95} for types earlier than M0 and from the young
disk populations described by \citet{leg92} for types of M0 and later.
The IR colors from \citet{kh95} are transformed from the
Johnson-Glass photometric system to the CIT system \citep{bb88}.
Near-IR colors in the 2MASS and CIT photometric systems agree at a level of
$<0.1$~mag \citep{car01}.

To estimate the amount of extinction toward each member of $\eta$~Cha, 
\citet{luh01} compared the $V-I$ color from \citet{law01} to the value 
for a dwarf at the published spectral type of that source. The resulting
extinctions ranged from $A_V=0$ to 0.8. However, when we include 
$V-R$ and $R-I$ in this analysis and use the spectral types measured in
this work, we arrive at extinctions that are close to zero for most of the
members. A comparison of our spectra to the estimates of unreddened 
spectra of young stars from \citet{luh04} also indicates essentially
no extinction, with a firm limit of $A_V<0.5$ for most members. 
Therefore, we take $A_V=0$ when computing bolometric luminosities from the 
observed photometry. The one exception is source 13 (HD~75505), whose 
optical colors do imply a non-zero extinction of $A_V=0.4$.

\begin{figure}
\plotone{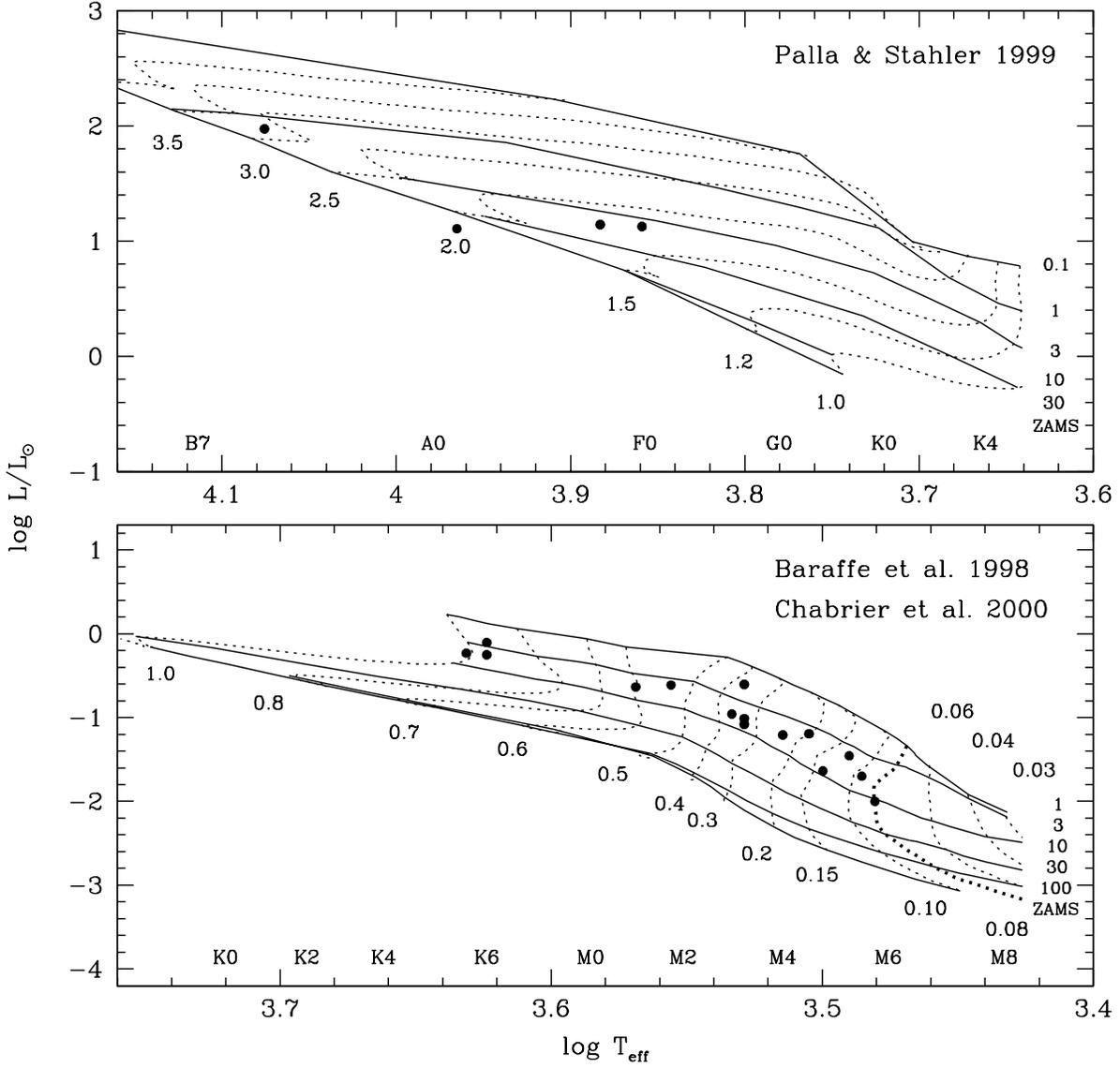}
\caption{
H-R diagram for the known members of the $\eta$~Cha young association.
These data are shown with the theoretical evolutionary models of \citet{pal99} 
({\it upper panel}) and \citet{bar98} ($0.1<M/M_\odot\leq1$) and \citet{cha00} 
($M/M_\odot\leq0.1$) ({\it lower panel}), where the mass tracks 
({\it dotted lines}) and isochrones ({\it solid lines}) are labeled in units 
of $M_\odot$ and Myr, respectively. These models imply an age of 
$6^{+2}_{-1}$~Myr for the association.
}
\label{fig:hr}
\end{figure}

Spectral types of M0 and earlier are converted to effective temperatures
with the dwarf temperature scale of \citet{sk82}.
For spectral types later than M0, we use the temperature scale that was designed
by \citet{luh03b} to be compatible with the models of \citet{bar98} and
\citet{cha00}. Bolometric luminosities are estimated by combining 
$J$-band measurements, a distance of 97~pc, and bolometric corrections 
described in \citet{luh99}.
The combined uncertainties in $A_J$, $J$, and BC$_J$ ($\sigma\sim0.14$, 0.03,
0.1) correspond to errors of $\pm0.07$ in the relative values of
log~$L_{\rm bol}$.
When an uncertainty in the distance modulus is included ($\sigma\sim0.1$),
the total uncertainties are $\pm0.12$. 
The temperatures and luminosities of the components of the binary system 
source 8 (RS Cha) are from \citet{rib00} and \citet{mam00}, respectively.
Two additional known binaries among the members of $\eta$~Cha are 
sources 1 and 9 \citep{kp02}.
The luminosities plotted on the H-R diagram in Figure~\ref{fig:hr} 
are corrected for the contribution of the secondary by assuming that for
each system the ratio of the luminosities is equal to the ratio the $K$-band 
fluxes from \citet{kp02}. Meanwhile, the luminosities in Table~\ref{tab:mem}
are uncorrected and reflect the total system photometry.
The effective temperatures, bolometric luminosities, and adopted
spectral types for the members of $\eta$~Cha are listed in Table~\ref{tab:mem}.

The temperatures and luminosities for the members of
$\eta$~Cha can be interpreted in terms of masses and ages with theoretical
evolutionary models. After considering the available sets of models,
\citet{luh03b} concluded that those of \citet{pal99} for $M/M_\odot>1$ and
\citet{bar98} and \citet{cha00} for $M/M_\odot\leq1$ provided the best 
agreement with observational constraints. The known members of $\eta$~Cha 
are plotted with these models on the H-R diagram in Figure~\ref{fig:hr}.
One member, source 12, appears overluminous compared to the other sources and 
thus may be an unresolved binarity. Otherwise, the remaining members form a 
relatively narrow, well-defined sequence that is parallel to the model 
isochrones and is consistent with a single age of $6^{+2}_{-1}$~Myr 
for all members. 
This age agrees within the uncertainties with previous estimates by 
\citet{mam99} and \citet{law01}.
According to the H-R diagram and evolutionary models in Figure~\ref{fig:hr},
the three newly discovered members of $\eta$~Cha have masses of 
0.08-0.13~$M_{\odot}$, and thus are just above and straddling the hydrogen 
burning mass limit.

\subsection{Implications of Survey}

We now examine the implications of our survey of the 
$\eta$~Cha association. We have identified three new members just above
the hydrogen burning limit, but have found no objects at lower masses down to 
the completeness limit of 0.015~$M_{\odot}$. Is this result consistent with 
the mass functions
measured in younger, star-forming clusters? In Taurus and IC~348, the
ratio of the number of brown dwarfs at 0.02-0.08~$M_{\odot}$ to the
number of stars has been measured to be 0.13 \citep{bri02,luh03a,luh03b}.
If this ratio applies to 
$\eta$~Cha, then we would expect to find only 2.3 substellar members in this 
mass range. Therefore, the lack of any detected brown dwarfs does not
comprise a statistically significant difference from mass functions in 
these two populations.
Meanwhile, $\eta$~Cha does appear to differ at a modestly significant 
level from the Trapezium Cluster in Orion,
which exhibits a frequency of brown dwarfs that is about twice the value
found in Taurus and IC~348 \citep{luh00,hc00,mue02}. 
As illustrated in Figure~\ref{fig:map}, the
three members of $\eta$~Cha with the lowest masses do have the largest
angular separations from the center of the association, which is suggestive
of mass segregation and could indicate that
substellar members of the association are preferentially
located in the outskirts of the association and beyond our survey field.
The possibility of the formation of a low-mass halo in $\eta$~Cha 
through dynamical evolution was discussed by \citet{mam00}.
A survey like the one performed in this work for a larger area toward
$\eta$~Cha would provide a test of this scenario.
Finally, we note that \citet{lyo04} predicted the presence of at least 20
undiscovered members at masses of 0.025-0.15~$M_{\odot}$ based on an
extrapolation of the IMF of the known members. However, we demonstrate
that only one new member beyond the census of \citet{lyo04} is present in 
this mass range.

\section{MBM~12A}

\subsection{Selection of Candidate Members}
\label{sec:ident2}

In a recent search for members of MBM~12A, \citet{ogu03} identified four 
objects as probable members of the association through slitless grism 
spectroscopy. Two of these sources were in the spectroscopic sample of 
\citet{luh01} and the other two stars are included in our sample here.
At the end of the survey for new members of MBM~12A by \citet{luh01}, 
there remained six candidates that had not been observed spectroscopically.
In this work, we selected the three brightest candidates for spectroscopy,
2MASS~J02524907+1952523, J02585238+1958402, and J02583445+1946506.
To facilitate the classification of the spectra of the candidates, we also
observed the known member MBM~12A-3.
Table~\ref{tab:log2} summarizes our long-slit spectroscopy of these targets.

\subsection{Classification of Candidate Members}
\label{sec:class2}

\citet{ogu03} obtained slitless grism spectroscopy toward the MBM~12 cloud
and identified four stars as ``probable" members of MBM~12A on the basis of
emission in H$\alpha$. 
However, H$\alpha$ emission also occurs in field dwarfs and thus can serve 
as evidence of youth and membership only if it is above the levels observed
in active dwarfs.
For instance, the equivalent widths from \citet{ogu03} for three of 
the four sources are less than 10~\AA, which is within the range observed
for field M dwarfs \citep{giz02}. For H$\alpha$ in the fourth star, they report
an equivalent width of 56.4~\AA\ while we measure only 10~\AA. 
We suggest that the measurement from \citet{ogu03} is either spurious or
obtained during a flare. In summary, the H$\alpha$ strengths for these four 
sources are consistent with both field dwarfs and young members of MBM~12A,
and hence cannot be used to assess membership in the association.

\begin{figure} 
\plotone{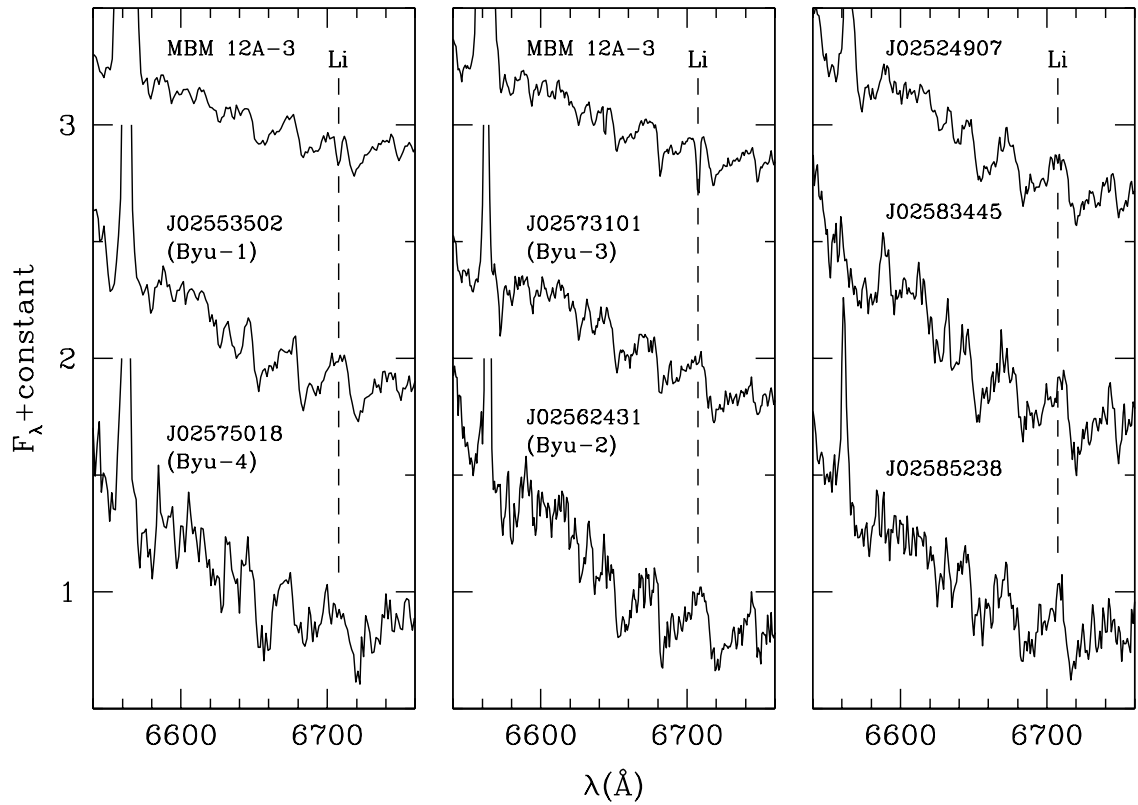} 
\caption{
Medium-resolution spectra of MBM~12A-3 (LkH$\alpha$~263), the four candidate 
members of MBM~12A presented by \citet{ogu03} ({\it left and middle}), and 
three candidate members from \citet{luh01} ({\it right}). 
The strong absorption in Li 6707~\AA\ that is found in young stars like
MBM~12A-3 is absent in the data for the candidates, 
indicating that they are field dwarfs rather than members of the association.
From left to right, the spectra have resolutions of 3.1, 2.1, and 2.8~\AA, 
respectively, and are normalized to the continuum near the Li line.}
\label{fig:li}
\end{figure}

To conclusively determine the membership of the four objects from \citet{ogu03},
we checked for the presence of strong Li absorption at 6707~\AA, 
which is a signature of youth.
Prior to the study of \citet{ogu03}, two of the four 
candidates therein, 2MASS~J02553502+2006484 (Byu-1) and 2MASS~J02575018+1958302
(Byu-4), already had been classified as field dwarfs through spectroscopy of 
Li by \citet{luh01}. Those data exhibit a clear absence of Li absorption,
as shown in Figure~\ref{fig:li}, where a spectrum of the known member 
MBM~12A-3 obtained with the same instrument configuration is included 
for comparison. Similarly, Li absorption is not present in the spectra 
obtained in this work for the other two sources from \citet{ogu03}, 
2MASS~J02562431+2035117 (Byu-2) and 2MASS~J02573101+1943459 (Byu-3).
Membership in MBM~12A for these two stars would have been somewhat surprising 
given their location below the boundary in the $R-I$ versus $I$ diagram 
from \citet{luh01}, which is the reason they were not observed 
spectroscopically or listed as candidates in that work.

\citet{luh01} found that the optical and near-IR colors of 
2MASS~J02524907+1952523, J02583445+1946506, and J02585238+1958402
were consistent with those of unreddened stars at spectral types of M4 to M5,
either foreground dwarfs or young members of MBM~12A.
From the spectra collected for these objects, we have measured spectral types
of M3.5V, M4.75V, and M4.75-M5V, which are in agreement with the estimates from 
the photometry. Given the lack of Li absorption in the spectra 
in Figure~\ref{fig:li}, we classify these stars as field dwarfs.

\section{Conclusions}

We have performed spectroscopic studies of candidate members of the 
$\eta$~Cha and MBM~12A young associations, the conclusions for which are 
summarized as follows:

For $\eta$~Cha, we have used $JHK_s$ photometry from 2MASS and
$i$ photometry from DENIS to construct color-color and color-magnitude 
diagrams for an area of 0.7~deg$^2$. Through spectroscopy of the candidate
members appearing in these data, we have discovered three new members
of the association with spectral types of M5.25-M5.75, corresponding 
to masses of 0.13-0.08~$M_{\odot}$ according to evolutionary models.
These sources were independently found in recent work by \citet{sz04} 
and \citet{lyo04}. No brown dwarfs were detected in $\eta$~Cha down to the 
completeness limit of 0.015~$M_{\odot}$ for our survey, which is roughly
consistent with the yield of $\sim2$ substellar members expected if the
relative numbers of stars and brown dwarfs in $\eta$~Cha are similar to those
of the Taurus and IC~348 star-forming regions. 
On the other hand, the fact that the three 
least massive members of the association are the outermost members may 
indicate that the substellar members are preferentially located at large
distances from the center of the association and outside of our survey field.
For the 18 known members of $\eta$~Cha, we have estimated bolometric 
luminosities and effective temperatures and placed the members on the
H-R diagram, from which we infer an age of $6^{+2}_{-1}$~Myr for the 
association with the evolutionary models of \citet{bar98}.

For MBM~12A, we have presented spectra of three of the remaining candidate 
members that lacked spectroscopy at the end of the survey by \citet{luh01}
and of the four ``probable" members from \citet{ogu03}. We classify all
of these sources as field dwarfs based on the absence of Li absorption in
their spectra.

\acknowledgements

We thank the staff at Las Campanas Observatory for their outstanding support
of these observations.
We are grateful to Isabelle Baraffe and Francesco Palla for access to their
most recent calculations. K. L. was supported by grant NAG5-11627 from the 
NASA Longterm Space Astrophysics program and D. S. acknowledges a Smithsonian 
Astrophysical Observatory Clay Fellowship.
2MASS is a joint project of the University of Massachusetts 
and the Infrared Processing and Analysis Center/California Institute 
of Technology, funded by the National Aeronautics and Space
Administration and the National Science Foundation.
DENIS is funded by the SCIENCE and the Human Capital and Mobility plans of
the European Commission under grants CT920791 and CT940627 in France,
by l'Institut National des Sciences de l'Univers, the Minist\`{e}re de
l'\'{E}ducation Nationale and the Centre National de la Recherche Scientifique 
(CNRS) in France, by the State of Baden-W\"{u}rtemberg in Germany, by the 
DGICYT in Spain, by the Sterrewacht Leiden in Holland, by the Consiglio 
Nazionale delle Ricerche (CNR) in Italy, by the Fonds zur F\"{o}rderung der 
wissenschaftlichen Forschung and Bundesministerium f\"{u}r Wissenschaft 
und Forschung in Austria, and by the ESO C \& EE grant A-04-046.

\clearpage

\begin{deluxetable}{llllll}
\tabletypesize{\scriptsize}
\tablewidth{0pt}
\tablecaption{Observing Log for $\eta$ Cha\label{tab:log1}}
\tablehead{
\colhead{} &
\colhead{} &
\colhead{Grating} &
\colhead{Resolution} &
\colhead{} \\
\colhead{Date} &
\colhead{Telescope + Instrument} &
\colhead{l~mm$^{-1}$} &
\colhead{\AA} &
\colhead{ID\tablenotemark{a}}
}
\startdata
2003 May 1 & Magellan~I + B\&C & 600 & 5 & 14,15,16,17,J08381023-7842153,J08481929-7911034,J08483486-7853513 \\
 & & & & J08362667-7902260,J08492628-7900397 \\
2003 May 2 & Magellan~I + B\&C & 600 & 5 & 3,4,5,6,9,10,12,13 \\
2003 June 24 & Magellan~I + B\&C & 1200 & 2.6 & 1,7,11,J08430440-7837463,J08423433-7832282,J08493005-7844228, \\
 & & & & J08453547-7843596,J08491452-7848397,J08411327-7910044, \\
 & & & & J08435227-7910133,J08440502-7855289 \\
\enddata
\tablenotetext{a}{Identifications are from Table~\ref{tab:mem} for members of 
$\eta$~Cha and from the 2MASS Point Source Catalog for the remaining stars.}
\end{deluxetable}

\begin{deluxetable}{lllllllllll}
\tabletypesize{\scriptsize}
\tablewidth{0pt}
\tablecaption{Field Stars Toward $\eta$ Cha \label{tab:back}}
\tablehead{
\colhead{} &
\colhead{} &
\colhead{} &
\colhead{} &
\colhead{} &
\colhead{Field Star} &
\colhead{} &
\colhead{} &
\colhead{} &
\colhead{} &
\colhead{} \\
\colhead{2MASS} &
\colhead{$\alpha$(J2000)\tablenotemark{a}} &
\colhead{$\delta$(J2000)\tablenotemark{a}} &
\colhead{Spectral Type} &
\colhead{Ref} &
\colhead{Evidence\tablenotemark{b}} &
\colhead{Ref} &
\colhead{$i$\tablenotemark{c}} &
\colhead{$J-H$\tablenotemark{a}} & \colhead{$H-K_s$\tablenotemark{a}}
& \colhead{$K_s$\tablenotemark{a}} 
}
\startdata
J08362667-7902260 &   08 36 26.67 &   -79 02 26.1 &                G8III &        1 &      sp &      1 &     9.53 &     0.71 &     0.17 &     7.59  \\
J08364076-7854583 &   08 36 40.77 &   -78 54 58.3 &                M2III &      2 & Li,sp,$\mu$ &    2 &    10.46 &     0.92 &     0.25 &     8.05  \\
J08375143-7844370 &   08 37 51.44 &   -78 44 37.1 &                K5III &      2 & Li,sp,$\mu$ &    2 &     9.81 &     0.83 &     0.17 &     7.69  \\
J08381023-7842153 &   08 38 10.23 &   -78 42 15.4 &                K2III &        1 &      sp &      1 &    10.11 &     0.58 &     0.18 &     8.42  \\
J08411327-7910044 &   08 41 13.28 &   -79 10 04.5 &                K2III &        1 &   Li,sp &      1 &     9.22 &     0.70 &     0.20 &     7.16  \\
J08423433-7832282 &   08 42 34.34 &   -78 32 28.2 &                K5III &        1 &   Li,sp &      1 &    11.14 &     0.80 &     0.22 &     8.99  \\
J08430440-7837463 &   08 43 04.41 &   -78 37 46.3 &                K4III &        1 &   Li,sp &      1 &    10.83 &     0.72 &     0.27 &     8.71  \\
J08435227-7910133 &   08 43 52.28 &   -79 10 13.4 &                K2III &        1 &   Li,sp &      1 &    10.42 &     0.61 &     0.18 &     8.68  \\
J08440502-7855289 &   08 44 05.03 &   -78 55 28.9 &                K1III &        1 &   Li,sp &      1 &     8.93 &     0.52 &     0.18 &     7.23  \\
J08453547-7843596 &   08 45 35.47 &   -78 43 59.7 &                K4III &        1 &   Li,sp &      1 &    11.62 &     0.77 &     0.22 &     9.44  \\
J08481929-7911034 &   08 48 19.29 &   -79 11 03.4 &                K2III &        1 &      sp &      1 &    10.48 &     0.60 &     0.15 &     8.75  \\
J08483486-7853513 &   08 48 34.87 &   -78 53 51.4 &           M5V,M4.75V &  2,1 & Li,$\mu$,NaK & 2,2,1 &    14.08 &     0.54 &     0.31 &    11.38  \\
J08491452-7848397 &   08 49 14.53 &   -78 48 39.7 &             K3-K4III &        1 &   Li,sp &      1 &    11.56 &     0.81 &     0.19 &     9.52  \\
J08492628-7900397 &   08 49 26.29 &   -79 00 39.7 &                K2III &        1 &      sp &      1 &    10.44 &     0.67 &     0.15 &     8.70  \\
J08493005-7844228 &   08 49 30.05 &   -78 44 22.8 &                K2III &        1 &   Li,sp &      1 &    11.64 &     0.80 &     0.19 &     9.51  \\
J08512320-7905232 &   08 51 23.20 &   -79 05 23.2 &                K4III &      2 & Li,sp,$\mu$ &    2 &     9.78 &     0.67 &     0.20 &     7.77  \\
\enddata
\tablecomments{Units of right ascension are hours, minutes, and seconds, and 
units of declination are degrees, arcminutes, and arcseconds.}
\tablenotetext{a}{2MASS Point Source Catalog.}
\tablenotetext{b}{Status as a field star is indicated by
a spectral classification as a dwarf or a giant (``sp"), 
weak Li absorption (``Li"), strong Na~I and K~I absorption (``NaK"),
or a proper motion that differs from that of
the known members of $\eta$ Cha (``$\mu$").}
\tablenotetext{c}{Second DENIS Release.}
\tablerefs{
(1) this work;
(2) \citet{law02}.}
\end{deluxetable}

\begin{deluxetable}{llllllllllllllllll}
\setlength{\tabcolsep}{0.04in}
\tabletypesize{\tiny}
\rotate
\tablewidth{0pt}
\tablecaption{Data for Members of $\eta$ Cha \label{tab:mem}}
\tablehead{
\colhead{} &
\colhead{Other} &
\colhead{} &
\colhead{} &
\colhead{} &
\colhead{} &
\colhead{} &
\colhead{Membership} &
\colhead{} &
\colhead{} &
\colhead{} &
\colhead{} &
\colhead{} &
\colhead{} &
\colhead{} &
\colhead{} &
\colhead{} &
\colhead{} \\
\colhead{ID\tablenotemark{a}} & 
\colhead{Names} &
\colhead{$\alpha$(J2000)\tablenotemark{b}} &
\colhead{$\delta$(J2000)\tablenotemark{b}} &
\colhead{Spectral Type\tablenotemark{c}} &
\colhead{Ref} &
\colhead{Adopt} &
\colhead{Evidence\tablenotemark{d}} &
\colhead{Ref} &
\colhead{$T_{\rm eff}$\tablenotemark{e}} &
\colhead{$L_{\rm bol}$} &
\colhead{$V-R$\tablenotemark{f}} & \colhead{$R-I$\tablenotemark{f}} &
\colhead{$I$\tablenotemark{f}} &
\colhead{$i$\tablenotemark{g}} &
\colhead{$J-H$\tablenotemark{b}} & \colhead{$H-K_s$\tablenotemark{b}} &
\colhead{$K_s$\tablenotemark{b}} 
}
\startdata
 1 & GSC 9402-0921 &   08 36 56.23 &   -78 56 45.5 & K4,K4,K6 & 1,2,3 &       K6 &    $\mu$,Li & 2,(1,2) &     4205 &      1.0 &     0.71 &     0.71 &     9.19 &     9.17 &     0.66 &     0.16 &     7.34  \\
 2 &   $\eta$ Cha &   08 41 19.48 &   -78 57 48.1 &       B8 & 4,2 &       B8 &  $\mu$,$\pi$ &      2 &    11900 &       94 &       \nodata &       \nodata &       \nodata &       \nodata &    -0.03 &     0.00 &     5.72  \\
 3 &       \nodata &   08 41 37.03 &   -79 03 30.4 & M3,M3.25 &  2,3 &    M3.25 &   Li,NaK &  2,3 &     3379 &    0.097 &     1.14 &     1.40 &    11.81 &    11.70 &     0.70 &     0.23 &     9.41  \\
 4 & GSC 9403-1083 &   08 42 23.73 &   -79 04 03.0 & K7,M1.75 &  2,3 &    M1.75 &       Li &      2 &     3596 &     0.24 &     0.92 &     1.07 &    10.79 &    10.71 &     0.76 &     0.16 &     8.62  \\
 5 &       \nodata &   08 42 27.11 &   -78 57 47.9 &    M5,M4 &  2,3 &       M4 &   Li,NaK &  2,3 &     3270 &    0.062 &     1.29 &     1.52 &    12.39 &    12.26 &     0.68 &     0.24 &     9.86  \\
 6 & GSC 9403-0288 &   08 42 38.80 &   -78 54 42.8 &    M2,M3 &  2,3 &       M3 &   Li,NaK &  2,3 &     3415 &     0.11 &     1.00 &     1.40 &    11.68 &    11.64 &     0.65 &     0.29 &     9.29  \\
 7 & Anonymous &   08 43 07.24 &   -79 04 52.5 & K4,K3,K6 & 1,2,3 &       K6 &       Li &  1,2 &     4205 &     0.79 &     0.67 &     0.73 &     9.44 &     9.40 &     0.66 &     0.12 &     7.63  \\
8A &  RS Cha A &   08 43 12.23 &   -79 04 12.3 &       A7 &      2 &       A7 &  $\mu$,$\pi$ &      2 &     7638 &     13.9 &       \nodata &       \nodata &       \nodata &       \nodata &     0.12 &     0.02 &     5.85  \\
8B & RS Cha B &       \nodata &       \nodata &  \nodata &  \nodata &  \nodata &  \nodata &  \nodata &     7228 &     13.4 &  \nodata &  \nodata &  \nodata &  \nodata &  \nodata &  \nodata &  \nodata  \\
 9 &       \nodata &   08 44 16.38 &   -78 59 08.1 &  M4,M4.5 &  2,3 &     M4.5 & Li,e,$\mu$,NaK & 2,2,5,3 &     3198 &    0.096 &     1.36 &     1.63 &    12.01 &    11.98 &     0.59 &     0.33 &     9.34  \\
10 & GSC 9403-1279 &   08 44 31.88 &   -78 46 31.2 & K7-M0,K7,M1 & 1,2,3 &       M1 &    Li,$\mu$ & (1,2),5 &     3705 &     0.23 &     0.86 &     0.92 &    10.75 &    10.71 &     0.73 &     0.19 &     8.73  \\
11 & GSC 9403-1016 &   08 47 01.66 &   -78 59 34.5 &  K4,K5.5 &  2,3 &     K5.5 &    Li,$\mu$ & 2,5 &     4278 &     0.59 &     0.66 &     0.71 &     9.76 &     9.84 &     0.70 &     0.37 &     7.66  \\
12 & GSC 9403-0389 &   08 47 56.77 &   -78 54 53.2 & M2,M2,M3.25 & 1,2,3 &    M3.25 & Li,$\mu$,NaK & (1,2),5,3 &     3379 &     0.25 &     1.07 &     1.30 &    10.80 &    10.83 &     0.64 &     0.27 &     8.41  \\
13 &  HD 75505 &   08 41 44.72 &   -79 02 53.1 &       A1 &     4 &       A1 &       $\mu$ &      2 &     9230 &       13 &     0.06 &     0.09 &     7.12 &       \nodata &     0.07 &     0.06 &     6.93  \\
14 & USNO Anon 1 &   08 41 30.30 &   -78 53 06.5 & M4,M4.75 & 6,3 &    M4.75 &   Li,NaK & 6,3 &     3161 &    0.023 &     1.52 &     1.73 &    13.82 &    13.58 &     0.57 &     0.26 &    10.98  \\
15 & USNO Anon 2 &   08 43 18.58 &   -79 05 18.2 & M2-M3,M3.25 & 6,3 &    M3.25 & Li,e,NaK & 6,(6,3),3 &     3379 &    0.083 &     0.99 &     1.21 &    11.77 &    11.81 &     0.67 &     0.40 &     9.43  \\
16 &       \nodata &   08 44 09.15 &   -78 33 45.7 & M4.5-M5.5,M5.75 &   7,3 &    M5.75 & Li,e,NaK & 7,(7,3),3 &     3024 &    0.010 &       \nodata &       \nodata &       \nodata &    14.79 &     0.53 &     0.36 &    11.62  \\
17 &       \nodata &   08 38 51.50 &   -79 16 13.7 & M5,M5.2,M5.25 & 7,8,3 &    M5.25 &   Li,NaK & (7,8),3 &     3091 &    0.035 &     1.61 &     1.93 &    13.28 &    13.28 &     0.55 &     0.29 &    10.43  \\
18 &       \nodata &   08 36 10.73 &   -79 08 18.4 &     M5.5 & 7,8,3 &     M5.5 &   Li,NaK & (7,8),3 &     3058 &    0.020 &     1.72 &     2.04 &    13.90 &    13.93 &     0.57 &     0.33 &    10.95  \\
\enddata
\tablecomments{Units of right ascension are hours, minutes, and seconds, and 
units of declination are degrees, arcminutes, and arcseconds.}
\tablenotetext{a}{1 through 12 are RECX designations from \citet{mam99}.
Identifications 13 through 18 are assigned in this work.}
\tablenotetext{b}{2MASS Point Source Catalog.}
\tablenotetext{c}{Measurement uncertainties for the spectral types from
this work are $\pm0.5$ and 0.25 subclass for K and M types, respectively.}
\tablenotetext{d}{Membership in $\eta$~Cha is indicated by 
strong emission in H$\alpha$ (``e"),
Na~I and K~I strengths intermediate between those of dwarfs and giants (``NaK"),
strong Li absorption (``Li"), 
or a proper motion (``$\mu$") or parallax (``$\pi$")
that is similar to that of the known members of $\eta$~Cha.}
\tablenotetext{e}{Temperature scale from \citet{sk82} ($\leq$M0) and
\citet{luh03b} ($>$M0), except for RS Cha A and B, where the temperatures
are from \citet{rib00}.}
\tablenotetext{f}{\citet{law01} and \citet{lyo04}.}
\tablenotetext{g}{Second DENIS Release.}
\tablerefs{
(1) \citet{cov97};
(2) \citet{mam99};
(3) this work;
(4) \citet{hc75};
(5) \citet{law01};
(6) \citet{law02};
(7) \citet{sz04};
(8) \citet{lyo04}.}
\end{deluxetable}

\begin{deluxetable}{llllll}
\tabletypesize{\scriptsize}
\tablewidth{0pt}
\tablecaption{Observing Log for MBM 12A\label{tab:log2}}
\tablehead{
\colhead{} &
\colhead{} &
\colhead{Grating} &
\colhead{Resolution} &
\colhead{} \\
\colhead{Date} &
\colhead{Telescope + Instrument} &
\colhead{l~mm$^{-1}$} &
\colhead{\AA} &
\colhead{ID\tablenotemark{a}}
}
\startdata
2002 January 12 & MMT + Blue Channel & 600 & 2.8 & J02524907+1952523,J02583445+1946506,J02585238+1958402 \\
2003 December 14 & MMT + Red Channel & 1200 & 2.1 & MBM~12A-3,J02562431+2035117,02573101+1943459 \\
\enddata
\tablenotetext{a}{Identifications are from the 2MASS Point Source Catalog,
except for MBM~12A-3, which is from \citet{luh01}.}
\end{deluxetable}

\clearpage

\clearpage

\clearpage

\clearpage

\clearpage

\clearpage

\clearpage


\newpage

\begin{thebibliography}{}


\bibitem[Andersson et al.(2002)]{and02}
Andersson, B.-G., Idzi, R., Uomoto, A., Wannier, P. G., Chen, B., \& 
Jorgensen, A. M. 2002, \aj, 124, 2164

\bibitem[Baraffe et al.(1998)]{bar98}
Baraffe, I., Chabrier, G., Allard, F., \& Hauschildt, P. H. 1998, \aap, 337, 403


\bibitem[Bessell \& Brett(1988)]{bb88}
Bessell, M. S., \& Brett, J. M. 1988, \pasp, 100, 1134


\bibitem[Brice\~no et al.(2002)]{bri02}
Brice\~{n}o, C., Luhman, K. L., Hartmann, L., Stauffer, J. R., \& Kirkpatrick, 
J. D. 2002, \apj, 580, 317

\bibitem[Burrows et al.(1997)]{bur97}
Burrows, A., et al. 1997, \apj, 491, 856


\bibitem[Carpenter(2001)]{car01}
Carpenter, J. M. 2001, \aj, 121, 2851

\bibitem[Chabrier et al.(2000)]{cha00}
Chabrier, G., Baraffe, I. Allard, F., \& Hauschildt, P. H. 2000,
\apj, 542, 464 

\bibitem[Covino et al.(1997)]{cov97}
Covino, E., Alcala, J. M., Allain, S., Bouvier, J., Terranegra, L., \&
Krautter, J. 1997, \aap, 328, 187

\bibitem[Gizis et al.(2000)]{giz00}
Gizis, J. E., Monet, D. G., Reid, I. N., Kirkpatrick, J. D., Liebert, J., 
\& Williams, R. J. 2000, \aj, 120, 1085

\bibitem[Gizis et al.(2002)]{giz02}
Gizis, J. E., Reid, I. N., \& Hawley, S. L. 2002, \aj, 123, 3356



\bibitem[Hearty et al.(2000)]{hea00}
Hearty, T., Fern\'{a}ndez, J. M., Alcal\'{a}, J. M., Covino, E.,
\& Neuh\"{a}user, R. 2000, \aap, 357, 681

\bibitem[Herbig \& Bell(1988)]{hb88}
Herbig, G. H., \& Bell, K. R. 1988, Lick Obs.\ Bull.\ Ser., No.\ 1111

\bibitem[Hillenbrand \& Carpenter(2000)]{hc00}
Hillenbrand, L. A., \& Carpenter, J. M. 2000, \apj, 540, 236

\bibitem[Hobbs et al.(1986)]{hbm86}
Hobbs, L. M., Blitz, L., \& Magnani, L. 1986, \apj, 306, L109

\bibitem[Houk \& Cowley(1975)]{hc75}
Houk, N., \& Cowley, A. P. 1975, University of Michigan Catalogue of Spectral
Types, Vol. 1., University of Michigan, Ann Arbor


\bibitem[Kenyon \& Hartmann(1995)]{kh95}
Kenyon, S. J., \& Hartmann, L. 1995, \apjs, 101, 117



\bibitem[K\"{o}hler \& Petr-Gotzens(2002)]{kp02}
K\"{o}hler, R., \& Petr-Gotzens, M. G. 2002, \aj, 124, 2899

\bibitem[Lawson et al.(2001)]{law01}
Lawson, W. A., Crause, L. A., Mamajek, E. E., \& Feigelson, E. D. 2001, \mnras, 
321, 57

\bibitem[Lawson et al.(2002)]{law02}
Lawson, W. A., Crause, L. A., Mamajek, E. E., \& Feigelson, E. D. 2002, \mnras, 
329, L29 

\bibitem[Leggett(1992)]{leg92}
Leggett, S. K. 1992, \apjs, 82, 351

\bibitem[Luhman(1999)]{luh99}
Luhman, K. L. 1999, \apj, 525, 466

\bibitem[Luhman(2001)]{luh01}
Luhman, K. L. 2001, \apj, 560, 287

\bibitem[Luhman(2004)]{luh04}
Luhman, K. L. 2004, \apj, 602, 816

\bibitem[Luhman et al.(2000)]{luh00}
Luhman, K. L., et al. 2000, \apj, 540, 1016

\bibitem[Luhman et al.(2003a)]{luh03a}
Luhman, K. L., Brice\~{n}o, C., Stauffer, J. R., Hartmann, L.,
Barrado y Navascu\'{e}s, D., \& Nelson, C. 2003a, \apj, 590, 348

\bibitem[Luhman et al.(2003b)]{luh03b}
Luhman, K. L., Stauffer, J. R., Muench, A. A., Rieke, G. H., Lada, E. A., 
Bouvier, J., \& Lada, C. J. 2003b, \apj, 593, 1093

\bibitem[Lyo et al.(2004)]{lyo04}
Lyo, A.-R, Lawson, W. A., Feigelson, E. D., \& Crause, L. A. 2004, \mnras, 347,
246

\bibitem[Magnani et al.(1985)]{mbm85}
Magnani, L., Blitz, L., \& Mundy, L. 1985, \apj, 295, 402

\bibitem[Mamajek et al.(1999)]{mam99}
Mamajek, E., Lawson, W. A., \& Feigelson, E. D. 1999, \apj, 516, 77

\bibitem[Mamajek et al.(2000)]{mam00}
Mamajek, E., Lawson, W. A., \& Feigelson, E. D. 2000, \apj, 544, 356


\bibitem[Mart{\'\i}n et al.(1996)]{mar96}
Mart{\'\i}n, E. L., Rebolo, R., \& Zapatero Osorio, M. R. 1996, \apj, 469, 706

\bibitem[Muench et al.(2002)]{mue02}
Muench, A. A., Lada, E. A., Lada, C. J., \& Alves, J. 2002, \apj, 573, 366

\bibitem[Ogura et al.(2003)]{ogu03}
Ogura, K., Sugitani, K., Magakian, T. Yu., Movsessian, T. A., Nikogossian, 
E. H., Itoh, Y., \& Tamura, M. 2003, \pasj, 55, L49

\bibitem[Palla \& Stahler(1999)]{pal99}
Palla, F., \& Stahler, S. W. 1999, \apj, 525, 772

\bibitem[Ribas et al.(2000)]{rib00}
Ribas, I., Jordi, C., Torra, J., \& Gim\'{e}nez, A. 2000, \mnras, 313, 99


\bibitem[Schmidt-Kaler(1982)]{sk82}
Schmidt-Kaler, T. 1982, in Landolt-Bornstein, Group VI, Vol. 2, ed. K.-H.
Hellwege (Berlin: Springer), 454

\bibitem[Song et al.(2004)]{sz04}
Song, I., Zuckerman, B., \& Bessell, M. S. 2004, \apj, 600, 1016




\bibitem[Strai\v{z}ys et al.(2002)]{str02}
Strai\v{z}ys, V., \v{C}ernis, K., Kazlauskas, \& Laugalys, V. 2002, 
Baltic Astronomy, 11, 231

\end{thebibliography}
\end{document}